\documentclass[aps,prl,twocolumn,showpacs,showkeys,superscriptaddress,nofootinbib,nobibnotes]{revtex4}
\newenvironment{frontmatter}{}{\maketitle}%

\usepackage{amsfonts,amssymb,bbm}

\newcommand{\reals}{\mathbbm{R}}
\newcommand{\pc}{\mathbbm{P}}
\newcommand{\F}{\mathcal{F}}

\def\b{}%
\def\e{}%
\def\new#1{#1}%

\begin{document}

\begin{frontmatter}

\title{Invariant length scale in relativistic kinematics
-- \\ Lessons from Dirichlet branes}

\author{Frederic P.~Schuller}
\email{F.P.Schuller@damtp.cam.ac.uk}
\affiliation{Perimeter Institute for Theoretical Physics, 35 King Street N, Waterloo, Ontario, N2J 2W9, Canada}
\affiliation{DAMTP, Wilberforce Road, Cambridge CB3 0WA, United Kingdom}
\author{Hendryk Pfeiffer}
\email{hpfeiffer@perimeterinstitute.ca}
\affiliation{DAMTP, Wilberforce Road, Cambridge CB3 0WA, United Kingdom}
\affiliation{Emmanuel College, St Andrew's Street, Cambridge CB2 3AP, United Kingdom}
\affiliation{Perimeter Institute for Theoretical Physics, 35 King Street N, Waterloo, Ontario, N2J 2W9, Canada}
\date{October 16, 2003}

\begin{abstract}

We show that Dirac--Born--Infeld theory possesses a hidden invariance
that enhances the local $O(1,p)$ Lorentz symmetry on a Dirichlet
$p$-brane to an $O(1,p) \times O(1,p)$ gauge group, encoding both an
invariant velocity and acceleration (or length) scale. \new{} This enlarged gauge group predicts 
consequences for the kinematics of
observers on Dirichlet branes, with admissible accelerations being
bounded from above. An important lesson \new{beyond string theory} is that a fundamental length
scale can be implemented into the kinematics of general
  relativity,
preserving both space-time as a smooth manifold
and local Lorentz symmetry, contrary to \new{common belief}. \new{We point out consequences for string phenomenology, classical gravity and atomic physics.}


\end{abstract}


\end{frontmatter}


Any candidate theory of quantum gravity must address the breakdown of
the classical smooth manifold picture of space-time at distances
comparable to the Planck length $\sqrt{\hbar G/c^3}$. The fundamental
structure of physical space-time is therefore expected to be crucially
different from smooth Lorentzian manifolds \cite{Kempf}. Some
proposals for quantum gravity theories address this issue {\sl ab
initio} by employing a discrete structure to describe space-time,
e.g., as a spin foam \cite{Perez}. Superstring theory, in contrast, is
formulated on conventional smooth ten-dimensional Lorentzian
space-time, but encodes a fundamental length scale $\ell$ in the
dynamics. Models \cite{braneworld} of the observable universe as a
four-dimensional Dirichlet brane propagating in this higher
dimensional manifold attract much attention in string
phenomenology. At the level of D-branes, Seiberg and Witten
\cite{Seiberg} observed a significant departure from the smooth
manifold picture of space-time, in accordance with our expectations
for quantum theories of gravity. They cast the D-brane dynamics at low
energy into  
a gauge theory on a non-commutative space-time. The latter encodes the
fundamental string length scale into the geometry without breaking
Lorentz invariance, but at the cost of giving up the smooth manifold
even in this low energy limit. The preservation of Lorentz invariance
is crucial, as this is one of the most accurately tested symmetries in
physics, with no indication that it is broken or deformed at any
observable energy scale \cite{Stecker:2003pw}.

In this Letter, we show that Dirichlet $p$-branes, considered as
smooth sub-manifolds of the string target space, possess a hidden
$O(1,p)$ invariance in the low energy limit. Remarkably, this
dynamical invariance can be absorbed into the \new{brane} geometry in a
way that preserves both the smooth manifold structure {\sl and} the
local Lorentz symmetry.  The resulting kinematics encode a maximal
acceleration \new{$\mathfrak{a}\sim 1/\ell^2$} in addition to the invariant
speed of light, by an extension, rather than deformation, of the local
Lorentz group $O(1,p)$ to $O(1,p)\times O(1,p)$. The strong
equivalence principle is not violated, since the inertial frames
coincide with those of general relativity. The derived geometry
provides the relevant kinematical framework for observers in
brane-world scenarios, with direct implications for string
phenomenology. 
While the technique devised in this Letter is motivated by the low
energy dynamics on D-branes, it can be applied independently of
its string theoretical origin in order to geometrically encode
the minimal length scale, expected in any theory of quantum
gravity, into the kinematics of generally relativistic theories,
preserving both local Lorentz symmetry and the strong principle of
equivalence. In the context of string phenomenology, our approach
describes the low energy dynamics on D-branes without giving up the
concept of a smooth space-time manifold. In the context of general
relativity, as we show below, it is related to the so-called
non-symmetric gravity theory which features a ghost-free
regularization of space-time singularities. Our formalism is finally
sufficiently manageable to connect to well-known experiments in atomic
physics. A correction to the Thomas precession \cite{BIKletter}
provides an experimental lower bound of $\mathfrak{a}\geq
10^{22}m/s^2$.

Let us now turn to the detailed demonstration of our claims.  For
technical simplicity, we consider a Dirichlet $p$-brane in bosonic
string theory, i.e., a $(p+1)$-dimensional Lorentzian {sub-manifold}
$(\Sigma,g)$ of a \new{26}-dimensional bulk space-time~\cite{GWG}, with a
gauge potential $A$ whose dynamics are governed by the
Dirac--Born--Infeld action
\begin{equation}
\label{BIorig}
  \int_\Sigma \ell^{-(p+1)}\sqrt{|\det g_{\mu\nu}|}\sqrt{\det({\delta^\mu}_\nu + {\F^\mu}_\nu)}
\end{equation}      
in the low energy limit~\cite{Callan}. The antisymmetric tensor
$\F_{\mu\nu} = B_{\mu\nu} + \ell^2 F_{\mu\nu}$ contains the pull-back
$B_{\mu\nu}$ of the Neveu-Schwarz two-form field to the brane, and the
field strength $F_{\mu\nu}$ of the Abelian gauge potential $A_\mu$ on
the brane $\Sigma$. Indices of fields living on the brane are raised
and lowered using the induced metric $g_{\mu\nu}$ on the brane. The
action (\ref{BIorig}) is manifestly invariant under local $O(1,p)$
transformations.  In order to exhibit an additional hidden $O(1,p)$
invariance, first observe that only even powers of $\F$ contribute to
the Lagrangian, as $\det(\delta + \F) = \exp \textrm{tr} \ln(1 + \F)$,
and the trace annihilates the odd powers of $\F$ in the series
expansion of the logarithm. We are therefore free to multiply the
field $\F$ by a unit square root $I$, i.e., $I^2=+1$, such that the
Lagrangian density of (\ref{BIorig}) can be written \b
\begin{equation}
\label{BIpc}
  \mathcal{L}_{\textrm{\tiny DBI}} = \ell^{-(p+1)}\sqrt{|\det g_{\mu\nu}|}\sqrt{\det({\delta^\mu}_\nu + I {\F^\mu}_\nu)}. 
\end{equation}
\e Rather than assuming that $I$ is real, it will turn out to be
enlightening to consider an algebraic extension $\pc$ of $\reals$,
\begin{equation}
  \pc := \{\, a+Ib\mid\, a,b\in\reals,\, I^2=+1\,\}
\end{equation}
with $I \in \pc$, but $I \not\in\reals$. The set $\pc$, equipped with
addition and multiplication inherited from $\reals$, is a commutative
ring with unit, which we term the {\sl pseudo-complex} numbers
\cite{Schuller:2002fn}. $\pc$ fails to be a \new{number} field due to the existence
of zero-divisors
\begin{equation}
  \pc^0 := \pc^0_+ \cup \pc^0_- = \langle\sigma_+\rangle_\reals \cup \langle\sigma_-\rangle_\reals,
\end{equation}
where $\sigma_\pm:=\frac{1}{2}(1\pm I)$. Note that the zero-divisors
$\pc^0$ do not have multiplicative inverses, and that $\sigma_\pm$ are
orthogonal projectors onto the eigenspaces of $I$,
$I\sigma_\pm=\pm\sigma_\pm$, $\sigma_\pm^2 = \sigma_\pm$ and $\sigma_+
\sigma_- = 0$. As we will show in the following, taking the
pseudo-complex form (\ref{BIpc}) of the Dirac--Born--Infeld Lagrangian
seriously directly leads to the \new{claimed} extension of relativistic
kinematics that incorporates a finite upper bound on
accelerations. Mathematically, the addition in (\ref{BIpc}) requires
the pseudo-complexification $(T_q\Sigma)_\pc$ of the tangent spaces
${T_q\Sigma}$ \cite{Moffat}, so that
\begin{equation}
\label{htensor}
  H^\mu{}_\nu:= {\delta^\mu}_\nu+ I {\F^\mu}_\nu
\end{equation}
is a well-defined tensor. A local frame at a point $q\in \Sigma$ of a
pseudo-complexified tangent space is specified by $(p+1)$ orthonormal
pseudo-complex vectors $E_a \in (T_q\Sigma)_\pc$ $\cong \pc^{p+1}$,
i.e., $g(E_a, E_b) = \eta_{ab}$,
where 
$g$ extends $\pc$-bilinearly and
$\eta$ is the $(p+1)$-dimensional Minkowski metric. On the dual
space of $(T_q\Sigma)_\pc$, we choose the dual basis $E^a$, such that $E^a
E_b = \delta^a_b$. There is a gauge freedom associated with the choice
of frame whose gauge group is the pseudo-complexified orthogonal group
of the Minkowski metric $\eta_{ab}$, i.e.,
\begin{equation}
\label{defrep}
  O_\pc(1,p) = \{\Lambda \in \textrm{End}(\pc^{p+1})\,\,|\, {\Lambda^m}_a {\Lambda^n}_b \eta _{mn} =  \eta_{ab}\},
\end{equation}
which, considered as a (real) Lie group, is of twice the dimension of
the real Lorentz group $O(1,p)$, and contains the latter as a
subgroup. Of direct physical interest are only classes of frames
related by transformations in the connection component of the identity
$SO^e_\pc$ of the special orthogonal group. Its vector representation
is easily obtained from the exponentiation of the matrices \b
${(M^{mn})^a}_b := \eta^{ma} \delta^n_b - \eta^{n a}
\delta^m_b$,
giving $\Lambda(\omega) = \exp(\omega_{mn} M^{mn})$,
with pseudo-complex parameters $\omega_{mn}\in\pc$. \e


In order to see how the Dirac--Born--Infeld tensor~(\ref{htensor})
transforms under pseudo-complexified Lorentz transformations, it is
necessary to study the representation theory of $SO^e_\pc(1,p)$ in
some detail.
First, we note from the definition (\ref{defrep}) that the
pseudo-complex Lorentz group as a Lie group decomposes into a direct
product of two real (proper orthochronous) Lorentz groups,
\begin{equation}
\label{splitting}
  SO^e_\pc(1,p) \cong SO^e(1,p) \times SO^e(1,p),
\end{equation}
by employing the zero-divisor decomposition $\Lambda =
\Lambda_+\sigma_+ + \Lambda_- \sigma_-$, where $\Lambda_+, \Lambda_-
\in SO^e(1,p)$. We can hence first apply the standard theory for
(real) Lie groups, then identify the pseudo-complex vector
representation $(T_q\Sigma)_\pc \cong \pc^{1+p}$, and finally study its
tensor powers.
The irreducible real representations $\mathcal{R}$ of $SO^e_\pc(1,p)$
are tensor products $V_1\otimes V_2$ of real representations $V_1,V_2$
of $SO(1,p)$, on which $\mathcal{R}(\Lambda)$ acts by
\begin{equation}
\label{irrep}
  \mathcal{R}(\Lambda)[v_1\otimes v_2] := \rho_1(\Lambda_+)[v_1] \otimes \rho_2(\Lambda_-)[v_2].
\end{equation}
It is convenient to denote the `small' representations of
$SO^e_\pc(1,p)$ by pairs $[d_1,d_2]$ where $d_j=\dim_\reals V_j$.
We define {\sl pseudo-complex conjugation} as the $\reals$-linear map
$*\colon\pc\to\pc$, sending $\sigma_\pm\mapsto\sigma_\mp$. 
The {\sl pseudo-complex conjugate\/} of a matrix is
$M^\dagger={(M^{T})}^*$, and $M$ is called {\sl pseudo-hermitean\/} if
$M^\dagger=M$.  For each representation $\mathcal{R}$, there exists a
not necessarily equivalent {\sl pseudo-complex conjugate\/}
representation $\mathcal{R}^*$ on the same space, defined by
$\mathcal{R}^*(\Lambda) := \mathcal{R}(\Lambda^*)$. 

It is well known that the `smallest' irreducible real representations
of $SO^e(1,p)$ are of dimension $d_t=1$ (trivial), $d_v=p+1$ (vector),
$d_a=p(p+1)/2$ (antisymmetric rank-$2$ tensor) and
$d_s=(p+1)(p+2)/2-1$ (traceless symmetric rank-$2$ tensor).
Considering direct sums of two irreducible pseudo-complex conjugate
representations, we obtain a large class of $\pc$-modules on which
$SO^e_\pc(1,p)$ is
represented. Clearly, the pseudo-complexified tangent spaces
$(T_q\Sigma)_\pc \cong \pc^{p+1}$ correspond to the {\sl vector
representation} $\mathcal{R}_v \cong [d_v,1]\oplus[1,d_v]$, whose
elements $v$ transform under $\Lambda\in SO^e_\pc(1,p)$ by
\begin{equation}
  v^m\mapsto \Lambda^m{}_n v^n=\sigma_+\Lambda_+^m{}_n v_+^n+\sigma_-\Lambda_-^m{}_n v_-^n,
\end{equation}
where the last expression is given in the zero-divisor decomposition
with $v=v_+\sigma_++v_-\sigma_-$, $v_\pm\in\reals^{p+1}$.
There is also a {\sl trivial} (scalar) representation $\mathcal{R}_t$
on the $\pc$-module
$\pc$,
a symmetric $\mathcal{R}_s$ and an antisymmetric second rank tensor
$\mathcal{R}_a$ with pseudo-complex coefficients.

Particularly interesting
is the {\sl mixed rank-$2$ tensor} representation
$\mathcal{R}_m:=[d_v,d_v]\oplus[d_v,d_v]^*$, whose elements are generic
rank-$2$ tensors $H=\sigma_+H_++\sigma_-H_-$ with pseudo-complex
coefficients and the mixed transformation rule
\begin{eqnarray}
\label{Htransform}
  H^{mn}&\mapsto& \Lambda^m{}_k{(\Lambda^\ast)}^n{}_l H^{kl}\nonumber\\
  &=& \sigma_+\Lambda_+^m{}_k\Lambda_-^n{}_l H_+^{kl}
  + \sigma_-\Lambda_-^m{}_k\Lambda_+^n{}_l H_-^{kl}.
\end{eqnarray}
As the transformation~(\ref{Htransform}) maps
pseudo-(anti-)her\-mitean tensors to pseudo-(anti-)hermitean
ones, this representation decomposes further into {\sl
pseudo-hermitean} $\mathcal{R}_H$ and {\sl pseudo-anti-hermitean
rank-$2$ tensors} $\mathcal{R}_{\overline H}$, both of real dimension
$d_v^2$.
Since the vector representation $\mathcal{R}_v$ is isomorphic to its
pseudo-complex conjugate $\mathcal{R}_v^*$, we can find all
irreducible second rank tensors of $SO_\pc^e(1,p)$ in the
decomposition,
\begin{equation}
\label{vectdecompose}
 \mathcal{R}_v\otimes_\reals\mathcal{R}_v \cong
  \mathcal{R}_t\oplus\mathcal{R}_a\oplus\mathcal{R}_s\oplus\mathcal{R}_H\oplus\mathcal{R}_{\overline H},
\end{equation}
where an $\reals$-bilinear tensor product must be used. 


The Dirac--Born--Infeld tensor~(\ref{htensor}) is pseudo-hermitean,
i.e., of type $\mathcal{R}_H$ and therefore transforms
as~(\ref{Htransform}). As $\mathcal{R}_H$ is irreducible, the
$\delta^\mu{}_\nu$- and the $\F^\mu{}_\nu$-part
of~(\ref{htensor}) will mix under generic pseudo-complex
transformations.  The space-time indices of $H_{\mu\nu}$
are hence converted to indices with respect to a local orthonormal
pseudo-complex basis by means of
\begin{equation}
\label{BItensorflat}
  H_{ab} := E^\mu_a E^{*\nu}_b H_{\mu\nu}.
\end{equation}
Note that the alternative choice,
$H_{ab} := E^{* \mu}_a E^\nu_b H_{\mu\nu}$,
presents a transition 
to the isomorphic conjugate representation $\mathcal{R}^*_H \cong
\mathcal{R}_H$, which physically corresponds to a charge conjugation,
as pseudo-complex conjugation $*$ acting on $H$ inverts the sign of
the electromagnetic field strength, which can be absorbed into a
redefinition of charges $q \mapsto -q$.
Observe that the Dirac--Born--Infeld Lagrangian (\ref{BIpc}) is charge
conjugation invariant. If the Dirac--Born--Infeld Lagrangian
(\ref{BIpc}) is expressed in terms of $H_{ab}$, \b
\begin{equation}
  \mathcal{L}_{\textrm{\tiny DBI}} = \ell^{-(p+1)}\sqrt{|\det g_{\mu\nu}|} \sqrt{(-1)^n \det E^a_\mu E^{*b}_\nu H_{ab}},
\end{equation}
\e it is manifestly invariant under \new{the} $SO^e_\pc(1,p)$ transformations
\begin{equation}
\label{Htransf}
  H_{ab} \mapsto {\Lambda^m}_a {\Lambda^{* n}}_b H_{mn},\qquad
  E_a \mapsto {\Lambda^m}_a E_m\label{Etransf}.
\end{equation}
Before we set out to reveal the physical interpretation of the
pseudo-complex frame changes,
we derive the most general form of pseudo-complex frames that can be
reached from real frames by means of $SO^e_\pc(1,p)$-transformations.
One can show that a pseudo-complex frame transformation $\Lambda \in
SO^e_\pc(1,p)$ uniquely decomposes into a product of a real Lorentz
transformation $L \in SO^e(1,p)$ and a pseudo-complex transformation
\new{$Q=\exp(\omega_{mn}M^{mn})$} with purely pseudo-imaginary
parameters \new{$\omega_{mn} = - \omega^*_{mn}$}. Therefore, the
most general pseudo-complex frame $E$ that can be reached from the
real frame $e$ is given by $E_a = {Q_a}^b {L_b}^c e_c$.
But ${L_b}^c e_c$ simply presents a change of the real frame, so that
it suffices to study the case where $L$ is the identity.  Now the
transformation $Q$ can be uniquely decomposed into a real and
imaginary part, such that
\begin{equation}
  E_a = {\gamma_a}^b (\delta_b^c + \frac{I}{\mathfrak{a}} 
  {\theta_b}^c) e_c,\label{Edecomp}
\end{equation}    
where $\mathfrak{a}$ is a parameter of inverse length dimension, to be
determined below. Using that $Q^* = Q^{-1}$, one easily finds
that
\begin{equation}
  \gamma_{ab} = \gamma_{ba} \quad \textrm{and} \quad \theta_{ab} = - \theta_{ba}.
\end{equation} 
A subsequent real Lorentz transformation clearly preserves these
symmetries, so that the restriction to $L=\textrm{id}$ was without
loss of generality. Note that $\theta$ determines $\gamma$ up to a
real Lorentz transformation, because $Q$ is invertible and hence
injective when acting as a linear map.


In order to unravel the physical meaning of the pseudo-complex frame
transformations (\ref{Htransf}),
consider a solution of $(1+3)$-dimensional Born--Infeld
electrodynamics, i.e.\ assume a flat background $\eta$, vanishing NS
two-form $B_{\mu\nu}=0$, and an electromagnetic field strength tensor
$F$ that, up to a real Lorentz transform $L$ is given by $F_{01} =
-F_{10} = E$, $F_{23} = -F_{32} = B$, while all other components
vanish. Acting on $H = \eta + I \ell^2 F$ with the transformation
\begin{equation}
  Q = \left(\begin{array}{cccc}\cosh(I\alpha) & \sinh(I\alpha) & 0 & 0\\
                           \sinh(I\alpha) & \cosh(I\alpha) & 0 & 0\\
                           0 & 0 & \cos(I\varphi) & \sin(I\varphi)\\
                           0 & 0 & -\sin(I\varphi) & \cos(I\varphi) 
\end{array}\right)\label{crittrafo}
\end{equation}
according to (\ref{Htransf}), one achieves the reduction of the
generically pseudo-complex Born--Infeld tensor $H$ to its 'metric'
real part
\begin{eqnarray}
  H'_{00} = -H'_{11} &=& 1/\cosh(2\alpha),\\ 
  H'_{22} = H'_{33} &=& -1/\cos(2\varphi),
\end{eqnarray} 
if one chooses the transformation parameters $I\alpha$ and $I\varphi$
such that
\begin{eqnarray}
  B &=& \ell^{-2} \tan(2 \varphi),\label{Bcond}\\ 
  E &=& \ell^{-2} \tanh(2 \alpha)\label{Econd}.
\end{eqnarray}
Note that whereas (\ref{Bcond}) always has a unique solution, the
condition (\ref{Econd}) is only soluble for $E < \ell^{-2}$, as is the
case for solutions of Born--Infeld electrodynamics \cite{Born}
with vanishing magnetic field.  Action of (\ref{crittrafo}) on $E_a$
according to (\ref{Etransf}) yields
\begin{equation}
  \gamma_{00} = -\gamma_{11} = \cosh(\alpha),\quad 
  \gamma_{22} = \gamma_{33} = -\cos(\varphi),
\end{equation}
and
\begin{equation}
  \theta_{01} = -\theta_{10} = \mathfrak{a} \tanh(\alpha),\label{transacc}\quad
  \theta_{23} = \theta_{32} = \mathfrak{a} \tan(\varphi)\label{transvel}.
\end{equation}
So far, we have discussed the $SO^e_\pc(1,p)$-invariance of Born--Infeld
electrodynamics. We now show how particles minimally coupled to the
gauge field $A$ inherit \new{$SO^e_\pc(1,p)$ as their kinematical group}. 
It is well known that the Lorentz force on a particle of rest mass $m$
and electric charge $q$ can be expressed in the coordinates of a local
orthonormal real frame $e_a$ by $m {\Omega_0}^b = q \eta^{bm} F_{m0}$,
where $\Omega_{ab} = - \Omega_{ba}$ is the Frenet--Serret tensor
associated with a test particle of four-velocity $u=e_0$ and attached
spatial frame $e_\alpha$, so that $\nabla_u e_a = {\Omega_a}^b e_b$.
The components $\Omega_{0\alpha}$ encode the 3-acceleration of the
particle world-line, and
$\epsilon_{\alpha\beta\gamma}\Omega_{\beta\gamma}$ is the angular
velocity of the spatial frame $e_\alpha$ with respect to a
Fermi--Walker transported frame.

We now give the pseudo-complex frame (\ref{Edecomp}) a physical
meaning by identifying the antisymmetric real Lorentz tensor $\theta$
with the Frenet--Serret tensor $\Omega$, so that the scalar
acceleration of a test particle, according to (\ref{transacc}), is
given by $\mathfrak{a} \tanh(\alpha)$, for some $\alpha \in
\reals$. This identification implies the modified Lorentz force law
\begin{equation}
  {m_a}^n {\Omega_n}^b = {q^b}_n {H'}^{nm} F_{ma},\label{modLorentz}
\end{equation}
where the rest mass $m$ and charge $q$ are corrected by
acceleration-dependent matrix-valued factors $\gamma = \gamma(\Omega)$
\begin{equation}
  {m_a}^n(\Omega) = m {\gamma_a}^n,\label{mmass}\quad
  {q_a}^n(\Omega) = q {{\gamma^{-1}}_a}^n\label{mcharge}, 
\end{equation}
and the magnetic field in the co-moving frame is used to determine the
motion of the attached spatial frame. Note that by applying the
pseudo-complex transformation $Q$ of~(\ref{crittrafo}), we have passed
to a {\sl co-accelerated\/} frame which is determined up to a local
real Lorentz transformation, given by the remaining freedom in
$\gamma$ once $\theta$ has been fixed. The modified Lorentz force
law~(\ref{modLorentz}) holds in this frame so that indeed all
quantities in~(\ref{modLorentz}) are real.
For small accelerations $\tanh(\alpha) \ll 1$, (\ref{modLorentz})
smoothly reduces to the standard relativistic Lorentz force law, if we set the so far undetermined parameter
$\mathfrak{a} := 2q/m\ell^2$. This correspondence for low accelerations
justifies the identification $\theta=\Omega$. Note that if the proper
acceleration tends to $\mathfrak{a}$, or the angular velocity of the
spatial frame to infinity, the dynamical charge (\ref{mcharge}) of the
test particle tends to zero. This switching-off of the charge is the
dynamical explanation of why the test particle is not further
accelerated by the electromagnetic field if it achieves the maximum
acceleration.

The kinematical meaning of the pseudo-complex Lorentz transformations
is now evident from (\ref{transacc}):
A rotation in the spatial $\alpha\beta$--plane with purely
pseudo-imaginary parameter $I\varphi$ effects a transformation
to a uniformly rotating frame with angular velocity $\mathfrak{a}
\tan(\varphi)$. By means of local pseudo-imaginary rotations, one can
therefore always arrange for non-rotating observers. Starting from
such a Fermi--Walker transported observer, a boost transformation with
purely pseudo-imaginary parameter $I\alpha$ effects a transformation
to a translationally accelerated frame of scalar acceleration
$\mathfrak{a} \tanh(\alpha)$, obviously respecting the invariant
acceleration scale $\mathfrak{a}$ as an upper limit.
The action of the real Lorentz group on inertial, i.e., real, frames
is clearly as in standard relativity. In particular, the inertial
observers of Dirac--Born--Infeld kinematics agree with the inertial
observers of standard general relativity. Thus, there is no
violation of the equivalence principle implied by the presence of a
length scale in relativity. Compared to standard relativity, we impose
one additional condition on admissible observers, namely that their
co-accelerated, i.e., pseudo-complex, frames are continuously
connected to inertial ones by $SO^e_\pc(1,p)$, thus encoding the
maximal acceleration scale.
The term {\sl sub-maximally accelerated\/} is justified because
whenever the local frame of such an admissible observer is
Fermi--Walker transported, then its scalar acceleration is indeed
bounded by $\mathfrak{a}$.
Such relativistic kinematics with an invariant length scale
can be considered independently of \new{their} origin in Dirac--Born--Infeld
theory. We have therefore kinematically implemented 
a maximal acceleration, an idea going  
back to Caianiello~\cite{Caia}, but 
here derived from the low energy
dynamics of D-branes.\\
There is an intriguing connection to Moffat's non-symmetric
gravitational theory \cite{Moffat2}. 
Using a generalized metric $g_{(\mu\nu)} + I g_{[\mu\nu]}$, he
formulates a 
theory which is free of ghost poles, tachyons and problems with
asymptotic boundary conditions. A static spherically symmetric
solution does not contain a black hole event horizon, so that the
information loss problem is resolved at the classical level. His
generalized metric is pseudo-hermitean \cite{Moffat}, and hence we now
recognize that it transforms under the same representation
$\mathcal{R}_H$ as the Dirac--Born--Infeld tensor.  This strongly
suggests that Moffat found a gravity theory of Born--Infeld type,
whose solutions indeed feature a regulation of gravitational
singularities.



\end{document}